\RequirePackage{amsmath}
\RequirePackage{fix-cm}
\documentclass[smallextended]{svjour3}       
\smartqed  

\usepackage{multirow}
\usepackage{graphicx}
\usepackage{natbib}
\usepackage{marvosym}
\usepackage{epstopdf}
\usepackage{makecell}
\usepackage{color}
\usepackage{colortbl}
\usepackage[table]{xcolor}
\usepackage{amsmath}
\usepackage{pifont}
\usepackage{amsfonts}
\usepackage{bm}
\usepackage{url}            
\usepackage{subfigure}
\usepackage{mathptmx}
\usepackage{algorithm}
\usepackage{algorithmic}
\usepackage[algo2e]{algorithm2e}

\usepackage{enumitem}
\setlist{leftmargin=8mm}

\newtheorem{assumption}{Assumption}

\newcommand{\xmark}{\ding{55}}
\usepackage{array}
\newcolumntype{L}[1]{>{\raggedright\let\newline\\\arraybackslash\hspace{0pt}}m{#1}}
\newcolumntype{C}[1]{>{\centering\let\newline  \\\arraybackslash\hspace{0pt}}m{#1}}
\newcolumntype{R}[1]{>{\raggedleft\let\newline \\\arraybackslash\hspace{0pt}}m{#1}}
\begin{document}

\title{Millionaire: A Hint-guided Approach for Crowdsourcing}

\titlerunning{Millionaire: A Hint-guided Approach for Crowdsourcing}

\author{Bo Han*        \and
        Quanming Yao*   \and
           Yuangang Pan     \and
           Ivor W. Tsang\(^{\dagger}\) \and
           Xiaokui Xiao \and
           Qiang Yang \and Masashi Sugiyama}

\institute{
\(*\) indicates equal contributions.\\
\(\dagger\) indicates the corresponding author.\\
Bo Han \at
Centre for Artificial Intelligence (CAI), University of Technology Sydney, Australia\\ \& Center for Advanced Intelligence Project, RIKEN, Japan\\
\email{bo.han@student.uts.edu.au}
\and
Quanming Yao \at
4Paradigm Inc., Beijing, China\\
\email{yaoquanming@4paradigm.com}
\and
Yuangang Pan  \at
Centre for Artificial Intelligence (CAI), University of Technology Sydney, Australia\\
\email{yuangang.pan@student.uts.edu.au}
\and
Ivor W. Tsang \at
Centre for Artificial Intelligence (CAI), University of Technology Sydney, Australia \\
\email{Ivor.Tsang@uts.edu.au}
\and
Xiaokui Xiao \at
Department of Computer Science, National University of Singapore, Singapore \\
\email{xkxiao@nus.edu.sg}
\and
Qiang Yang \at
Department of Computer Science and Engineering, Hong Kong University of Science and Technology, Hong Kong \\
\email{qyang@cse.ust.hk}
\and
Masashi Sugiyama \at
Center for Advanced Intelligence Project, RIKEN, Japan \& Graduate School of Frontier Sciences, The University of Tokyo, Japan\\
\email{sugi@k.u-tokyo.ac.jp}
}

\date{Received: date / Accepted: date}

\maketitle
\begin{abstract}
Modern machine learning is migrating to the era of complex models, which requires a plethora of well-annotated data.
While crowdsourcing is a promising tool to achieve this goal,
existing crowdsourcing approaches barely acquire a sufficient amount of high-quality labels.
In this paper,
motivated by the ``Guess-with-Hints'' answer strategy from the Millionaire game show, we introduce the hint-guided approach into crowdsourcing to deal with this challenge. Our approach encourages workers to get help from hints when they are unsure of questions. Specifically, we propose a hybrid-stage setting, consisting of the main stage and the hint stage. When workers face any uncertain question on the main stage, they are allowed to enter the hint stage and look up hints before making any answer. A unique payment mechanism that meets two important design principles for crowdsourcing is developed. Besides, the proposed mechanism further encourages high-quality workers less using hints, which helps identify and assigns larger possible payment to them.
Experiments are performed on Amazon Mechanical Turk,
which show that our approach ensures a sufficient number of high-quality labels with low expenditure and detects high-quality workers.

\end{abstract}

\section{Introduction}
\label{intro}

Huge and complex models are popularly used in today's machine learning applications,
since they can take advantage of big data to get better performance.
Indeed,
they have significantly boosted performance of many important tasks,
such as image classification \citep{russakovsky2015imagenet},
speech recognition \citep{hinton2012deep},
dialogue systems \citep{sordoni2015}
and autonomous driving \citep{bojarski2016end}.
More recently,
they even beat a human champion by a large margin in the game Go \citep{silver2016mastering}.
However,
a primary question arises:
how can we provide sufficient fuel (a plethora
of annotated data) to propel our rocket (complex models)?
The most appealing way may
be the crowdsourcing technology \citep{russakovsky2015imagenet,zhong2015active,wang2016cost,wang2017obtaining},
since the process of annotations is convenient and the cost of annotations is very cheap.

While crowdsourcing techniques have been commonly used in many commercial platforms,
	such as Amazon Mechanical Turk (AMT),
	the quality of crowdsourced labels is not satisfactory \citep{ipeirotis2010quality}.
The reasons are that workers may not be domain experts \citep{vuurens2011much,yan2014learning,rodrigues2014sequence}. For example, it is hard for an average person to distinguish some professional tasks, such as labeling bird images or medical data \citep{wais2010}. Besides, some workers can just be spammers, who response questions with arbitrary answers \citep{difallah2012mechanical,raykar2012eliminating}. Such low-quality labels inevitably degenerates the performance of subsequent learning models \citep{natarajan2013learning,sukhbaatar2015training,han2016convergence}.
For instance, noisy labels degrade the accuracy of deep neural networks by 20\% to 40\% \citep{patrini2017making,yu2017transfer,yu2017learning}.

The previous efforts have extensively focused on statistical inferences,
which aggregate crowdsourced labels when they are already collected \citep{karger2011iterative,liu2012variational,chen2013optimistic,zhou2014aggregating,
tian2015max,zhang2016spectral}.
However, as crowdsourced labels are intrinsically noisy, statistical inferences are hard to guarantee that aggregated labels are reliable. In order to improve the label quality, recently, many researchers resort to a complementary direction, namely proposing approaches controlling the process of label collection \citep{singla2015incentivizing,litman2015relationship,
chen2016truthful,pennockbounded,zheng2015qasca,fan2015icrowd,han2017robust}.

These approaches aim to encourage workers to provide more reliable labels at the stage of collection. For example, the skip-based approach encourages workers to skip uncertain tasks. However, if many tasks are difficult, the label requester may collect only a few labels, which are not enough for subsequent learning models \citep{shah2015double,ding2016crowdsourcing}. The self-corrected approach encourages workers to check whether they need to correct their answers after looking at references. However, references consisting of responses from other workers are noisy, which may mislead workers.
Moreover, this approach is not realized on crowdsourcing platforms \citep{shah2016noopps}.
Therefore, existing approaches fail to acquire a sufficient number of high-quality labels on real tasks (Table~\ref{tab:comparison}).
Besides,
these approaches cannot detect and give potentially larger payment to high-quality workers.
However,
this is very important
as these workers are always preferred by crowdsourcing platforms,
thus they should be identified and more paid \citep{ipeirotis2010quality}.

\begin{table}[ht]
	\caption{Comparison of related approaches and our hint-guided approach (in bold).
		Baseline is the approach used by AMT.
		Note that, the self-corrected approach is designed theoretically, but barely realized for real tasks (denoted as ``\xmark''),
			so its metrics of ``label quality'' and ``money cost'' cannot be evaluated (denoted as ``-'').
			However,
			since its payment mechanism has a multiplicative form, it prevents spammers theoretically.}
	\centering
	\scalebox{0.85}{
		\begin{tabular}{c|c| c | C{65px} | C{70px} | C{50px}}
			\hline
			Perspective   &          Metric          &  Baseline & Skip-based~\citep{shah2015double,ding2016crowdsourcing} & Self-corrected~\citep{shah2016noopps} & \textbf{Hint-guided} (proposed) \\ \hline
			requester    &   large label quantity   & \checkmark & \xmark                           & \checkmark     & \checkmark           \\ \cline{2-6}
			&    high label quality    &   \xmark   & \checkmark                       & -              & \checkmark           \\ \hline
			worker     & quality detection &   \xmark   & \xmark                           & \xmark         & \checkmark           \\ \cline{2-6}
			&    spammer prevention    &   \xmark   & \checkmark                       & \checkmark     & \checkmark           \\ \hline
			platform    &      low money cost      &   \xmark   & \checkmark                       & -              & \checkmark           \\ \cline{2-6}
			&       realization        & \checkmark & \checkmark                       & \xmark         & \checkmark           \\ \hline
	\end{tabular}}
	\label{tab:comparison}
\end{table}

\begin{figure}[htp]
\centering
\subfigure[Main stage.]
{\includegraphics[width=0.85\textwidth]{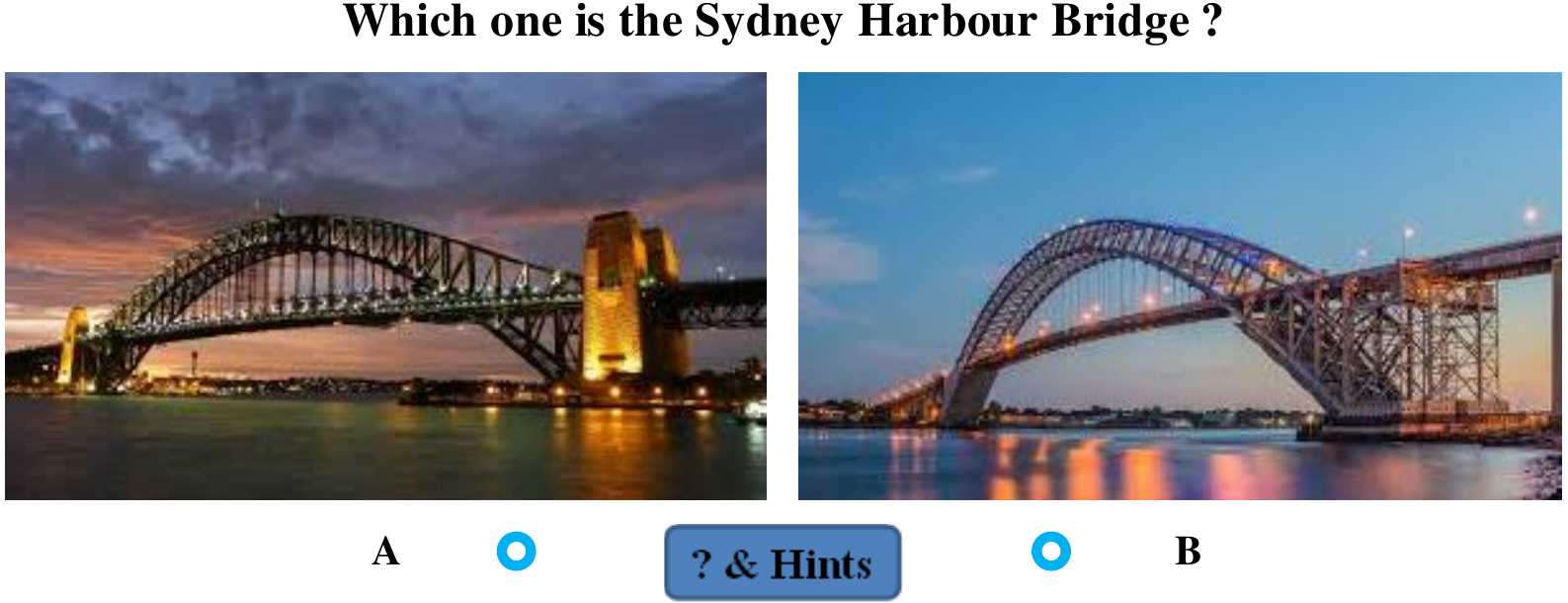}
	\label{fig:mainstage}}

\subfigure[Hint stage.]
{\includegraphics[width=0.85\textwidth]{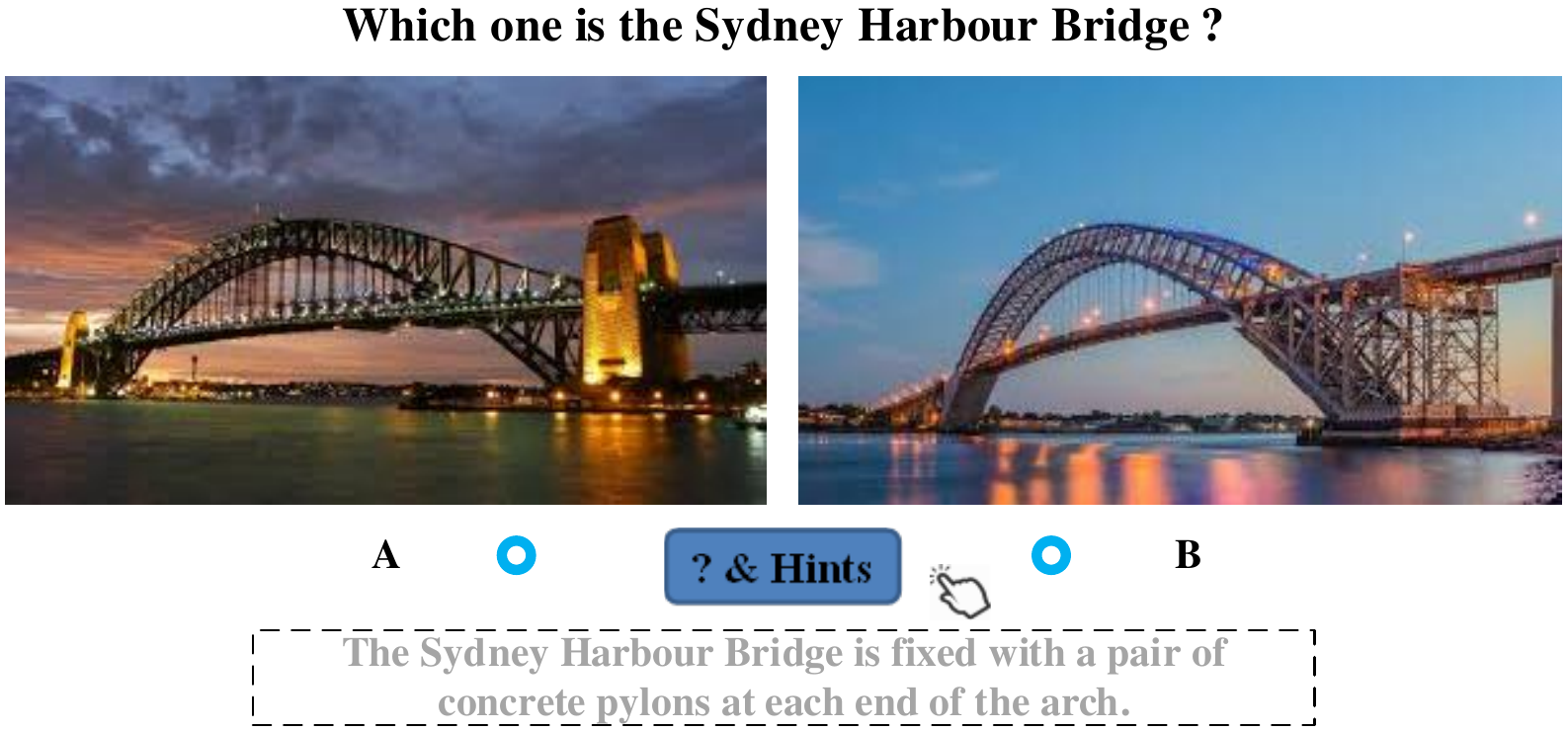}
	\label{fig:hintstage}}

\caption{A task that requires workers to answer the question ``Which one is the Sydney Harbour Bridge?". {\bf Top panel}: the proposed interface under the hybrid-stage setting, consists of two options (``A" and ``B") and a ``? \& Hints" button. {\bf Bottom panel}: when workers feel unsure of this question and click the button, the content of hints (gray) is visible, which guides workers to make a choice.}
\label{fig:interface}
\end{figure}

To address these issues, we are inspired by the ``Guess-with-Hints'' answer strategy from
the Millionaire
game show
\footnote{\url{https://en.wikipedia.org/wiki/Who_Wants_to_Be_a_Millionaire_(U.S._game_show)}},
where a challenger has opportunities to request hints from the show host when he/she feels unsure of the questions. By this strategy, we introduce a hint-guided approach to improve the quality of crowdsourced labels.
This approach encourages workers to get help from auxiliary hints when they answer questions that they are unsure of. To be specific, we introduce a hybrid-stage setting, which consists of the main stage and the hint stage. In the main stage,
for each question, workers answer it directly when they feel confident or jump into the hint stage when they feel uncertain. Once they enter the hint stage, they are allowed to look up hints before making any answer to this unsure question. The less number of times workers enter the hint stage, the higher quality they are estimated to be. To realize this setting, we provide an explicit ``? \& Hints" button (the bottom panel in Figure~\ref{fig:interface}) for each question. For example, when the worker is unsure of the question in Figure~\ref{fig:interface}, he/she can click this button and answer the question under the help of hints (the gray sentence).

Nevertheless, only hybrid-stage setting is not enough to address all issues. For example, if hints are freely available in the hint stage, even high-quality workers may abuse free hints for higher accuracy and rewards. This issue causes failure in the detection of high-quality workers.
Under the hybrid-stage setting, we develop a hint-guided payment mechanism,
which aims to incentivize workers to use the hints properly.
Specifically, our mechanism penalizes workers who use the hints.
Therefore, high-quality workers will answer most of the questions directly (without hints) for higher rewards.
Then, our mechanism assists our setting to detect the high-quality workers effectively.
Moreover, we prove that our mechanism is unique under the hybrid-stage setting.
Since our mechanism has a multiplicative form,
it prevents spammers as well. Our contributions are summarized as follows.

\begin{itemize}
\setlength\itemsep{5px}

\item In crowdsourcing, our hint-guided approach is the first attempt to improve the quality of labels by auxiliary hints, and detect the high-quality workers. Our approach is different in both setting and payment mechanism from existing approaches
such as self-corrected and skip-based approaches.

\item We introduce a hybrid-stage setting. Under this setting, we propose a hint-guided payment mechanism, which incentivizes workers to use hints properly instead of abusing them. Moreover, we prove the uniqueness of our mechanism under the proposed setting.

\item We further give some general rules for task requester,
	which helps them easily design hints for their own tasks.

\item We conduct comprehensive experiments on AMT platforms. Empirical results on three real tasks show that, the proposed approach reaches an excellent performance on the adequate collection of high-quality labels using low expenditure.
Meanwhile, our approach prevents spammers and detects high-quality workers as well.

\end{itemize}

The remainder of this paper is organized as follows. In Section \ref{related-literature}, related literature is presented. Section \ref{problem-setting} introduces the novel setup in crowdsourcing, namely the hybrid-stage setting. In Section \ref{sec:mechanism}, we propose a hint-guided payment mechanism under this setting. In Section \ref{Numerical-Experiments}, we provide the experiment setup and empirical results related to three real-world tasks. The conclusions are given in Section~\ref{Conclusions}.

\section{Related Literature}\label{related-literature}

\subsection{Post-Processed Approach}

In crowdsourcing, the statistical inference (post-processed) approach is popularly used to improve the quality of labels \citep{zhang2016learning}.
Such approach tries to find the correct label for each question only after noisy labels being collected from the platform.
Many methods have been developed under this approach.

For example, \citet{Raykar2010} presented a two-coin probabilistic model,
where each worker's labels are generated by flipping the ground-truth labels with a certain probability.
\citet{Yan2010} extended this two-coin model by setting the dynamic flipping probability associated with samples.
\citet{Kajino2012} formulated a probabilistic multi-task model, where each worker is considered as a task.
\citet{Zhou2012} proposed a minimax entropy model.
\citet{Bi2014} employed a mixture probabilistic model for worker annotations, which learns a prediction model directly.
\citet{tian2015max} extended weighted Majority Voting by the max-margin principle,
which provides a geometric interpretation of crowdsourcing margin.
However,
as labels are intrinsically noisy,
it is hard for this type of approach to obtain a sufficient among of correct labels with statistical inference.

\subsection{Pre-Processed Approach}

While previous efforts have extensively focused on several statistical inferences,
pre-processing approach has been recently developed as an alternative way to improve label quality.
Namely, the crowdsourced setting is coupled with the payment mechanism,
which incentivizes workers to provide more reliable labels at the stage of label collection.
Thus, unlike post-processed approach,
pre-processed one can directly reduce the noise in obtained labels.
Moreover,
post-processed approach can be used to further reduce the noise in labels
after they are obtained by the pre-processed approach.

In this paper,
we target the pre-processed approach from the perspective of machine learning \citep{buhrmester2011amazon, singla2013truthful,goel2014mechanism,ho2015incentivizing,lambert2015axiomatic,shah2015double, ding2016crowdsourcing}.
The most related works are the skip-based \citep{shah2015double, ding2016crowdsourcing} and self-corrected approaches \citep{shah2016noopps}.
In the skip-based approach,
workers are allowed to select a skip option based on their confidence for each question.
However, this in turn leads to insufficient label quantity.
A two-stage setting is used in the self-corrected approach.
Workers firstly answer all questions in the first stage,
and then they are allowed to correct their first-stage answers after looking at a reference in the second stage.
However, references consisting of responses from other workers are noisy, which may mislead workers to providing incorrect labels.
Besides,
as a reference needs to be set for each task, such a setting is not supported by the AMT platform and only simulation results are reported in \citet{shah2016noopps}.
Finally, neither the skip-based nor self-corrected approaches can identify worker quality as our approach.

The pre-processed approach was also considered in the database area, but the focus is different. Normally, their research is to dynamically assign the optimal $K$ ($ \leq N$) problems to each worker by his/her work quality,
where $N$ is the total number of problems to be annotated
\citep{zheng2015qasca,fan2015icrowd}. Thus, worker quality control plays an fundamental role in the quality of crowdsourcing from the viewpoint of database.

\subsection{Worker Quality Control}

As workers' quality has huge impact on the obtained labels,
	many researchers tried to improve label quality by offering better control over workers' quality.
For example,
\citet{raykar2012eliminating} considered detecting spammers or adversarial behavior, and tried to eliminate them in the following iterations or phases. However, this method does not consider how to detect high-quality workers.
Then,
\citet{joglekar2013evaluating} devised techniques to generate confidence intervals for worker error rate estimates, thereby enabling a better evaluation of worker quality. However, this method is complex to be deployed. For our hybrid-stage setting, the less number of times workers enter the hint stage, the higher quality they are estimated to be.

\section{Problem Setup}
\label{problem-setting}
Inspired by the ``Guess-with-Hints'' answer strategy,
we introduce the hint-guided approach to improve the quality of crowdsourced labels and detect the high-quality workers at the same time. This approach encourages workers to get help from the useful hints when they answer uncertain questions (Figure~\ref{fig:interface}). Specifically, we realize this approach in Section~\ref{sec:hint-guided-approach}, including the hybrid-stage setting and the payment mechanism. Then,
easy usage of hints is discussed in Section~\ref{sec:richhint}.
Finally,
the rationality of our design is discussed in Section~\ref{sec:needhint}.

\subsection{Hint-guided Approach}\label{sec:hint-guided-approach}
Here, we describe our hint-guided approach from the following four aspects.
\subsubsection{Hybrid-stage Setting}
\label{sec:hybridset}

We first set up definitions for the hybrid-stage setting that consists of the main stage and the hint stage.
To model our setting, let us consider a simple example:
each worker answers $N$ binary-valued (objective) questions,
and each question has precisely one correct answer, either ``A'' or ``B''.
Therefore, for every question $i \in \{1, \ldots, N\}$,
a worker chooses an answer matching his/her own belief under the following hybrid-stage setting.

\begin{itemize}
\setlength\itemsep{10px}
	
\item
The main stage (Figure~\ref{fig:mainstage}):
For question $i$, he/she should be incentivized to select the option that he/she feels confident.
When he/she feels unsure and clicks the ``? \& Hints" button,
he/she jumps into the hint stage formalized by the ``H" option, namely,
\begin{align*}
\text{select}
\;
\begin{cases}
``A"  & \text{if}\; P_{A,i} \in [\frac{1}{2} + \epsilon,1)\\
``B"  & \text{if}\; P_{A,i} \in (0,\frac{1}{2} - \epsilon]\\
``H"  & \text{otherwise}
\end{cases},
\end{align*}
where $\epsilon \in [0, \frac{1}{2})$ models the worker's uncertainty degree in this stage,
$P_{A,i}$ is the probability of the worker's belief that the answer to the $i$th question is ``A''
(i.e., the probability that the worker believes ``A'' is the correct answer for the $i$th question).

\item
The hint stage (Figure~\ref{fig:hintstage}):
When he/she feels unsure of the question,
the worker clicks the ``? \& Hints" button. This means that he/she enters the hint stage.
Then,
the worker picks up ``A'' or ``B'' according to

\begin{align*}
\text{select}
\;
\begin{cases}
``A"  & \text{if}\; P_{A|H,i} \in [T,1)\\
``B"  & \text{if}\; P_{B|H,i} \in [T,1)
\end{cases},
\end{align*}
where
$T \in (\frac{1}{2}, 1)$ is the predefined threshold value of the worker's belief in the hint stage, $P_{A|H,i}$ is the probability of the worker's belief that the answer to the $i$th question is ``A'' given hints,
and $P_{B|H,i}$ is the probability of the worker's belief that the answer to the $i$th question is ``B'' given hints ($P_{B|H,i} = 1 - P_{A|H,i}$).
\end{itemize}

\begin{figure}[H]
\centering
\includegraphics[width=0.55\textwidth]{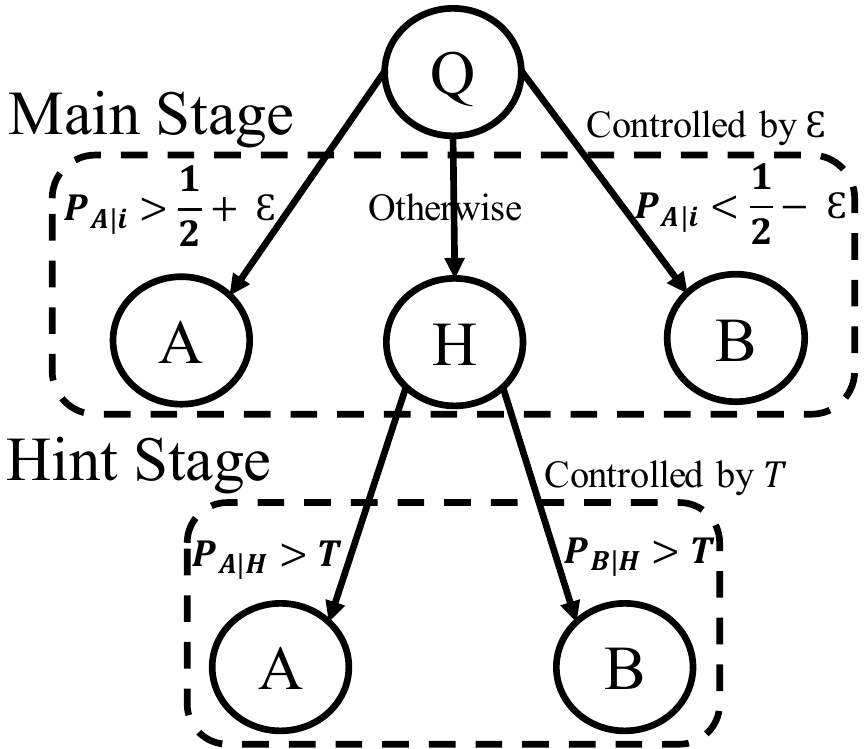}
\caption{Mathematical model of the decision process under our hybrid-stage setting.}
\label{fig:notation}
\end{figure}

The above modeling of the decision process is also summarized in Figure~\ref{fig:notation}.
As we can see,
$\epsilon$ controls the decision in the main stage
and the hint stage depends on $T$.
When $\epsilon$ is large, i.e., $\epsilon \rightarrow \frac{1}{2}$,
more workers need hints to make their decision for each question.
When $\epsilon$ is smaller, i.e., $\epsilon \rightarrow 0$,
fewer workers need hints to make their decision for each question.
Once the worker enters the hint stage,
when $T$ is set to a large value, i.e., $T \rightarrow 1$,
he/she will become more confident to make his/her final decision for each question.
When $T$ is set to a small value,
i.e., $T \rightarrow \frac{1}{2}$,
he/she will be less confident to make his/her final decision for each question.

Note that, $\epsilon$ is decided by $T$ according to Proposition~\ref{pr:epsilonlower} in Section~\ref{sec:IC-PM},
and $T$ is controlled by a mechanism designer. The choice of $T$ is based on different applications and given to us.
In the experiments,
we empirically choose $T = 0.75$ due to the qualitative psychology~\citep{smith2007qualitative}.

\subsubsection{Model Assumption}
Based on the hybrid-stage setting, we will introduce the corresponding payment mechanism, where it is rooted in the following assumption.

\begin{assumption}
	\label{ass:common}
	\begin{itemize}
		
		\item[(A).] There are $G$ ``gold standard" questions ($1 \leq G \leq N$),
		of which answers are known to the requester,
		uniformly distributed at random positions among all $N$ questions;
		
		\item[(B).] Each worker aims to maximize his/her expected payment for $N$ questions;
		
	\end{itemize}
\end{assumption}

Assumption~\ref{ass:common} is a standard one in analyzing pre-processed approaches for crowdsourcing
\citep{shah2015double,shah2016noopps,zhang2016learning}.
Specifically,
as answers to the ``gold standard" questions are known to the requester in advanced,
workers' responses to them can be used to evaluate workers' performance and decide payment for workers.
This is the functionality of Assumption~\ref{ass:common}~(A).
Then,
Assumption~\ref{ass:common}~(B) is a must for analyzing workers' performance.
It originates from game theory \citep{nisan2007algorithmic},
and means that each work wants to maximize its revenue.

Next,
we make the following Assumption~\ref{ass:extra},
which specifies our usage of hints here.
It is motivated by the educational psychology \citep{koedinger2007exploring},
and means that the hints are useful enough to guide workers to making final decisions.

\begin{assumption}
\label{ass:extra}
Workers have enough confidence to make a final decision after acquiring useful hints,
i.e., $T \in (\frac{5}{8}, 1)$ in the hint state.
\end{assumption}

Note that the confidence of a random guess is $T = \frac{1}{2}$,
thus $T > \frac{5}{8}$ means that the worker's confidence to pick up an answer is high after looking at the hint.
As an illustration,
	let us see Figure~\ref{fig:mainstage}.
	Workers outside Australia may not know which one is the Sydney Harbour Bridge.
	However, after reading the hints (grey) in Figure~\ref{fig:hintstage},
	workers should have enough confidence to make a final decision ``A''
	as the pylons structure is very obvious.
When $T$ approaches to $1$, the beliefs from the hint are maximal, or equivalently, the hint provides the worker with a certain answer.

\subsubsection{Payment Mechanism}
According to the model assumption, we are ready to introduce our payment mechanism based on the hybrid-stage setting. Specifically, after the worker answers all $N$ questions in the hybrid-stage setting,
his/her performance is evaluated by his/her responses to $G$ ($\leq N$) questions. Namely, his/her choice for each question in the gold standard gets evaluated to one of four states,
denoted by $\{\mathbb{D}_{+}, \mathbb{D}_{-}, \mathbb{H}_{+}, \mathbb{H}_{-}\}$. We define the four states as follows.
\begin{itemize}
\setlength{\itemsep}{5px}

\item $\mathbb{D}_{+}$: answer in the main stage and correct;

\item $\mathbb{D}_{-}$: answer in the main stage and incorrect;

\item $\mathbb{H}_{+}$: answer in the hint stage and correct;

\item $\mathbb{H}_{-}$: answer in the hint stage and incorrect.

\end{itemize}

Note that ``answer in the main stage" means that he/she feels confident in the main stage and answers directly;
``answer in the hint stage" means that he/she feels unsure in the main stage and answers with hints in the hint stage.
``correct" or ``incorrect" denotes whether the worker's selection matched with the standard answer in $G$ questions or not.

Therefore, under the hybrid-stage setting,
we can formulate any payment mechanism as function
\begin{align}
f: \left\lbrace
\mathbb{D}_{+}, \mathbb{D}_{-}, \mathbb{H}_{+}, \mathbb{H}_{-}
\right\rbrace^G \rightarrow [\mu_{\min}, \mu_{\max}],
\label{eq:scoref}
\end{align}
where $\min f(\cdot) = \mu_{\min}$ and $\max f(\cdot) = \mu_{\max}$. We reserve the rights to set $\mu_{\min}$ and $\mu_{\max}$, where $0 \leq \mu_{\min} \leq \mu_{\max}$. In our paper, the goal is to design $f$ such that its expected payment for each worker is strictly maximized under the above setting.

\subsubsection{Difference from Previous Approaches}

The most related approach to ours is the self-corrected approach \citep{shah2016noopps}, since both of us have two phases in the setting. However, they are totally different in probabilistic modeling. The self-corrected approach builds up the two-stage setting, and workers are necessarily required to enter the second stage to check the reference answer of every question,
whereas, our approach builds up the hybrid-stage setting, and workers are not necessary to enter the hint stage related to confident questions. Besides,
since each payment mechanism is customized based on
some designed goals (examples are in Section~\ref{sec:designing-rule})
under its corresponding setting,
our hint-guided payment mechanism is also different from the one used in the self-corrected approach.

It would be also interesting to discuss the advantages of the proposed approach over active learning~\citep{yan2011active} for crowdsourcing. There are two points to be highlighted. First, compared with active learning, hints in our approach may not be as strong as querying the ground-truth label; the hint only guides the worker to make a choice. Second, active learning is constrained to query which data sample should be labeled next and which annotator should be queried to benefit the learning model. However, our approach is free of these restrictions.

\subsection{General Rules of Hints}
\label{sec:richhint}

Motivated by instructional hints in the educational psychology~\citep{koedinger2007exploring},
to make the hints useful and reduce interface designers' workloads,
we offer three general rules here:
\begin{itemize}
\setlength{\itemsep}{3px}
	
\item [(A). ] The hints should be easily accessible to interface designers;

\item [(B). ] The hints should be discriminative and concise for workers; and

\item [(C). ] The hints should be irrelevant to the number of annotated samples in each task.
\end{itemize}

We adopt the three rules in designing our hints in experiments.
We take three practical datasets in our experimental setup (Section~\ref{sec:setup})
to justify that these requirements are reasonable in the real world.
First, for \textit{Sydney Bridge}, as an interface designer, we
easily acquire the content of hints from Wikipedia,
which includes discriminative and concise phrases, such as ``concrete pylons'' and ``around Sydney Opera House''.
Second, for \textit{Stanford Dogs}, we build a lookup table as hints,
which includes the characteristics of four breeds of dogs, such as “prick ears” for Norwich Terrier. It means that, the hints in this dataset should be irrelevant to the number of annotated samples, but relevant to the number of classes. Third, for \textit{Speech Clips}, the tool is freely available online to roughly recognize each speech clip and save the concise keywords ($\leq 4$) as the hints.


\subsection{Needs of Hybrid-stage Setting}
\label{sec:needhint}

It is also noted that, in designing the pre-processing mechanism, the high-quality worker detection is very important for collecting a sufficient number of high-quality labels. If the tasks can be assigned to each worker by his/her work quality, the annotation quality will be increased accordingly. Also, if we can detect the high-quality workers and give more weights on his/her annotations, we can acquire the better label aggregation.
Here,
we show that it may not be achieved by a single-stage setting with hints
(i.e., only Figure~\ref{fig:hintstage} and no Figure~\ref{fig:mainstage}).
Later,
we also empirically demonstrate this point in Section~\ref{exp:worker-quality-detection}.

Specifically,
by our Assumption~\ref{ass:extra},
if we want to collect more correct labels,
it is more naturally to directly assign visible hints for every single question.
This removes the necessity to have a hybird-stage setting as we design here.
However,
high-quality workers are always preferred by crowdsourcing platforms,
	thus they should be identified and more paid.
Such a fundamental goal may not be achieved by a simple single-stage setting with visible hints.
The reason is explained as follows.
Under the single-stage setting, both high-quality and low-quality workers can easily read the visible hints to answer questions.
Thus,
we cannot make a difference between them.
However, under the hybrid-stage setting, high-quality workers may not read the hints frequently.
Namely, the less number of times workers enter the hint stage, the higher quality they are estimated to be.
Thus, we can track the high-quality workers by our setting.

Note that, only this setting may encounter a problem:
if the hints are freely available in the hint stage, by Assumption~\ref{ass:common}~(B), even high-quality workers may abuse free hints for higher accuracy and rewards.
This issue causes failure in the detection of high-quality workers by the hybrid-stage setting.
Therefore, under the hybrid-stage setting, we hope to develop a payment mechanism (Section~\ref{sec:mechanism}),
which incentivizes workers to use the hints properly.
Specifically, this mechanism penalizes workers who use the hints.
Then, high-quality workers will answer most of the questions directly for higher rewards.
As a result, this mechanism helps our setting to detect the high-quality workers effectively.

\section{Hint-guided Payment Mechanism}
\label{sec:mechanism}

In Section~\ref{sec:designing-rule},
we first give two important definitions which help us to design a payment function.
Then,
the designed payment function is given in Section~\ref{sec:IC-PM}.
Furthermore,
we prove that our incentive-compatible payment mechanism is also unique under the hybrid-stage setting.
Finally,
in Section~\ref{sec:Harsh-axiom},
we clarify that more restrictive designing goals cannot be realized here.

\subsection{Design Principles}
\label{sec:designing-rule}

Incentive compatibility (Definition~\ref{def:incecomp}) and mild no-free-lunch axiom (Definition~\ref{def:axiom})
are important to design a payment mechanism for pre-processed approaches,
which are also popularly used by previous works \citep{shah2015double,shah2016noopps}.

\begin{definition}[Incentive Compatibility]
	\label{def:incecomp}
	A payment mechanism $f$ is incentive-compatible only if the following two conditions are satisfied:
	(i) $f$ gives an incentive to a worker to choose all answers by his/her belief;
	(ii) The expected payment, from the worker's belief, is strictly maximized in both the main stage and the hint stage.
\end{definition}

Definition~\ref{def:incecomp},
which is adapted from the standard game theoretical assumption \citep{nisan2007algorithmic},
describes incentive compatibility.
Basically,
it means that $f$ should encourage a worker to select the option he/she believes most likely to be correct.

\begin{definition}[Mild No-free-lunch Axiom]
\label{def:axiom}
If all answers attempted by a worker in ``gold standard" questions are either wrong or based on hints, then the payment for the worker should be zero, unless all answers attempted by the worker are correct. More formally, $f(\mathbf{a}) = 0$, $\forall \mathbf{a} \in \{
\mathbb{D}_{-}, \mathbb{H}_{+}, \mathbb{H}_{-}\}^G\backslash\{\mathbb{H}_{+}\}^G$.
\end{definition}

Definition~\ref{def:axiom} is a variant of the no-free-lunch axioms for our hybrid-stage setting.
It requires that $f$ should not pay a worker who has bad performance on ``gold standard'' questions.
This helps to reject spammers and keep high-quality workers, since answers to these questions are known to the platform and spammers are likely to give wrong answers while high-quality workers are not.

Our aim is to design the payment mechanism $f$, which is defined in Eq. \eqref{eq:scoref}, simultaneously satisfies the above definitions.

\subsection{Proposed Payment Mechanism}
\label{sec:IC-PM}

In order to design a payment mechanism,
we first consider the easiest case,
i.e., for a single question, how the worker should get paid under our hybrid-stage setting.
This helps us to find specific rules under Definition~\ref{def:incecomp} for our setting
under Assumption~\ref{ass:common},
such rules are given in Proposition~\ref{pr:conditions} below.

\begin{proposition}
\label{pr:conditions}

Let $f: \{\mathbb{D}_{+}, \mathbb{D}_{-}, \mathbb{H}_{+}, \mathbb{H}_{-}\} \rightarrow [0, \mu_{max}]$,
$d_+ = f\left( \mathbb{D}_{+} \right)$,
$d_- = f\left( \mathbb{D}_{-} \right)$,
$h_+ = f\left( \mathbb{H}_{+} \right)$
and $h_- = f(\mathbb{H}_{-})$.
When $N = G = 1$, $f$ satisfies Definition~\ref{def:incecomp} if it meets the following pricing constraints:

\begin{itemize}
\setlength{\itemsep}{1.5pt}
\item[(A). ]
$d_+ > d_-, h_+ > h_-, d_+ > h_+$.

\item[(B). ]
$\frac{d_+ - d_-}{1-2\epsilon} \geq \frac{h_+ - h_-}{2\epsilon}$.

\item[(C). ]
$d_+ - d_- \leq \frac{2T-1}{1/2 - \epsilon}\left( h_+ - h_- \right)$.

\end{itemize}
\end{proposition}

Condition (A) highlights that, for each question,
the payment $h_+$ from an indirect correct answer (after reading hints) should be less than $d_+$ from a direct correct answer.
Condition (B) bridges the per unit income gap $\frac{d_+ - d_-}{1-2\epsilon}$ in the main stage and $\frac{h_+ - h_-}{2\epsilon}$ in the hint stage together,
and the inequality encourages a worker to directly answer questions that he/she feels confident about in the main-stage.
Condition (C) incentivizes his/her to leverage the hints before answering questions that he/she is unsure of. Thus,
	conditions (A) and (B) encourage workers to directly answer questions without hints if they are confident enough;
	and when a worker has really low confidence,
	condition (C) encourages him/her to use hints.

\begin{remark}\it
In condition (A),
we cannot set $d_+ = h_+$.
If $d_+ = h_+$, even high-quality workers may abuse hints for higher accuracy and rewards.
This issue fails the detection of high-quality workers by the hybrid-stage setting.
Therefore, $d_+ > h_+$ ensures that high-quality workers will answer most of the questions directly for higher rewards.
Under the hybrid-stage setting, whether hints are used can be taken as a criterion to detect the high-quality workers.
Then, $d_+ > h_+$ assists our setting to detect the high-quality workers, which has been verified in experiments in Section~\ref{exp:worker-quality-detection}.
\end{remark}

From Proposition~\ref{pr:conditions},
we can see that $f$ relies on workers' uncertainty degree $\epsilon$ in the main stage and their confidence $T$ in the hint stage.
When $\epsilon$ is set to a large value, more workers need hints to make their decision for each question.
The disadvantage of large $\epsilon$ is that the overall payments for workers may be low due to leveraging too many hints.
When $\epsilon$ is set to a small value, fewer workers need hints to make their decision for each question.
The disadvantage of small $\epsilon$ is that the quality of crowdsourced labels may be poor since more workers avoid hints for higher payments.
Thus, we need to find $\epsilon$ to achieve a good tradeoff such that most workers are balanced, neither too cautious nor too careless.

However,
Proposition~\ref{pr:conditions} only makes use of Assumption~\ref{ass:common} to find rules for $f$
and does not specify the relationship between $\epsilon$ and $T$.
Below Proposition~\ref{pr:epsilonlower}
helps to connect $\epsilon$ and $T$,
and shows a lower-bound of $\epsilon$.

\begin{proposition}
\label{pr:epsilonlower}
Under Assumption~\ref{ass:common},
$f$ satisfies both Definitions~\ref{def:incecomp} and \ref{def:axiom} if $\epsilon \in [\epsilon_{\min}, 1/2)$
where $\epsilon_{\min} = T - \sqrt{T^2 - 1/4}$.
\end{proposition}

Moreover,
based on above Proposition,
we can derived following Corollary which is based on Assumption~\ref{ass:extra}.

\begin{corollary}
\label{cor:fh}
Under Assumption~\ref{ass:extra},
$(1/2 - \epsilon_{min}) < (2T - 1)$.
\end{corollary}



Finally,
we show when $\epsilon = \epsilon_{\min}$,
i.e.,
the boundary condition in Proposition~\ref{pr:epsilonlower} is achieved,
a hint-guided payment mechanism $f$ can be designed (Algorithm \ref{Hints-Payment}).
The function $g$,
which sets how a single question should be paid,
is defined at step~1 in Algorithm~\ref{Hints-Payment}.
Note that $g(\mathbb{H}_{+}) < g(\mathbb{D}_{+}) = 1$ due to Corollary~\ref{cor:fh},
which is also in consistent with condition (a) in Proposition~\ref{pr:conditions}.
Responses from workers on ``gold standard'' questions are collected in step~2,
and the budget is set in step~3.
A multiplicative form of $g$ is adopted in step 4,
which is inspired by \citet{shah2015double}.
It incentivizes workers to use hints properly and also helps to make the smallest payment to spammers.
The reasons are highlighted in Remark~\ref{rmk:multiplicative}.

\begin{algorithm}[ht]
\caption{Hint-guided Payment Mechanism\label{Hints-Payment}}
{\bfseries Inputs:}

\begin{itemize}
	
\item[1.]
Define a function $g: \{\mathbb{D}_{+}, \mathbb{D}_{-}, \mathbb{H}_{+}, \mathbb{H}_{-}\} \rightarrow \mathbb{R}_+$,
which sets how a single question should be paid, and ,
where $g(\mathbb{D}_{+}) = 1$,
$g(\mathbb{D}_{-}) = 0$,
$g(\mathbb{H}_{+}) = \frac{1/2 - \epsilon_{min}}{(2T - 1)}$
and $g(\mathbb{H}_{-})=0$;

\item[2.]
$a_1, \ldots, a_G \in \{\mathbb{D}_{+}, \mathbb{D}_{-}, \mathbb{H}_{+}, \mathbb{H}_{-}\}$ are the state evaluations of the answers to the $G$ gold standard questions;

\item[3.]
Set the minimum payment $\mu_{\min}$ and maximum payment $\mu_{\max}$ properly;
\end{itemize}

{\bfseries The payment is:}

\begin{itemize}
\item[4.]
$f\left( \left[ a_1, \ldots, a_G \right] \right) = \beta \prod_{i=1}^{G} g(a_{i}) + \mu_{\min}$ where $\beta = \mu_{\max} - \mu_{\min}$.
\end{itemize}
\end{algorithm}

\begin{remark}\it
\label{rmk:multiplicative}
The benefits of using the multiplicative form is detailed as follows.
For example,
a spammer will respond to a question with an arbitrary answer,
thus he/she will get the minimum payment once any answers in ``gold standard" are wrong.
Then,
for a normal worker,
if he/she tries to get the highest payment,
he/she is encouraged to use hints as less as possible.
The reason is that
the payment for a correct answer after using hints is $g(\mathbb{H}_{+})$ which is smaller than $1$,
i.e., $g(\mathbb{D}_+)$ (Corollary~\ref{cor:fh}).
Thus,
more hints are used,
the maximum payment for a worker will get smaller.
Besides,
such a multiplicative form also helps us to identify and pay more for high-quality workers,
as those workers will naturally user less hints.
\end{remark}

The design of Algorithm \ref{Hints-Payment} is further supported by the following Theorem~\ref{thm:one-mechanism}.
Thus, our algorithm is the unique one to satisfy both Definitions~\ref{def:incecomp} and \ref{def:axiom},
and $\epsilon = \epsilon_{\min}$ is also a must choice here. Note that, in practice,
the algorithm makes the minimum payment $\mu_{\min}$ instead of $0$ in Definition~\ref{def:axiom},
if one or more attempted answers in the gold standard are wrong.
This operation is without any loss of generality.

\begin{theorem}
\label{thm:one-mechanism}
Under Assumption~\ref{ass:common} and \ref{ass:extra},
$f$ in Algorithm \ref{Hints-Payment} satisfies both Definitions~\ref{def:incecomp} and \ref{def:axiom}
if and only if $\epsilon = \epsilon_{\min}$.
\end{theorem}

\subsection{No Other Compatible Mechanism}
\label{sec:Harsh-axiom}

Definition~\ref{def:incecomp} is a must to design a payment mechanism.
However,
under our hybird-setting here,
there exists another popular ``harsh no-free-lunch'' axiom (Definition~\ref{def:axiom-h}),
which is adapted from Definition 2 in~\citet{shah2016noopps}.

\begin{definition}[Harsh No-free-lunch Axiom]
\label{def:axiom-h}
If all answers attempted by the worker in ``gold standard" questions are either wrong or based on hints, then the payment for the worker should be zero. More formally, $f(\mathbf{a}) = 0$, $\mathbf{a} \in \{\mathbb{D}_{-}, \mathbb{H}_{+}, \mathbb{H}_{-}\}^G$.
\end{definition}

Compared to the ``mild no-free-lunch" axiom,
Definition~\ref{def:axiom-h} encourages the worker to answer without hints no matter whether he/she is unsure.
Thus,
it is stronger than the ``mild no-free-lunch'' axiom and can be used to replace Definition~\ref{def:axiom}. We wonder whether we can find another payment function which satisfies this more restrictive condition.
However,
below Theorem~\ref{thm:harsh} shows a contradiction to Definition \ref{def:axiom-h}.

\begin{theorem}
\label{thm:harsh}
Under Assumption~\ref{ass:common} and \ref{ass:extra},
there is no mechanism that satisfies both Definitions~\ref{def:incecomp} and \ref{def:axiom-h}.
\end{theorem}


Therefore,
the ``harsh no-free-lunch" axiom is too strong for the existence of any incentive-compatible payment mechanism here.
This further illustrates the uniqueness of the proposed payment mechanism.

\section{Numerical Experiments}\label{Numerical-Experiments}

We conduct real-world experiments on Amazon Mechanical Turk
\footnote{\url{https://www.mturk.com/mturk/welcome}},
which is the leading platform to collect crowdsourced labels.
We compare our hint-guided approach with:
(1) Baseline approach : a single-stage setting with an additive payment mechanism.
(2) Skip-based approach \citep{shah2015double,ding2016crowdsourcing}: a skip-stage setting with a skip-based payment mechanism. Note that the skip-based payment mechanism is multiplicative.
For the self-corrected approach \citep{shah2016noopps}, it has not been verified on AMT tasks, since there is no criteria how to set references. Therefore, we do not include it in our comparison. Note that additive and multiplicative payment mechanisms are respectively denoted as ``$+$" and ``$\times$" for subsequent use in Section~\ref{spammer-prevention}.

\subsection{Experimental Setup}
\label{sec:setup}

All these datasets are collected by us on Amazon MTurk, where hints are easily designed according to the criteria in Section~\ref{sec:richhint}. We conducted three real tasks as follows.
\begin{itemize}

\item \textit{Sydney Bridge} (binary-choice questions): we collect $30$ images of various bridges. Each image contains one bridge. The task is to identify whether the bridge in each image is the Sydney Bridge. The content of hints includes discriminative phrases, such as ``concrete pylons'' and ``around Sydney Opera House''.

\item \textit{Stanford Dogs} (multiple-choice questions): we collect $100$ images of four breeds of dogs. The task is to identify the breed of dogs in each image. We build a lookup table as hints, which includes the characteristics of four breeds of dogs, such as ``prick ears" for Norwich Terrier.

\item \textit{Speech Clips} (subjective questions): we collect $10$ speech clips. Each speech clip consists of $1$ or $2$ short sentences ($15$ words). The task is to recognize each speech clip and write down the corresponding sentence. We leverage the open tool
\footnote{\url{https://speech-to-text-demo.mybluemix.net/}} to roughly recognize each speech clip and save the key words ($\leq 4$) as the hints.

\end{itemize}

We verify the effectiveness of our hint-guided approach from three perspectives (Table~\ref{tab:comparison}),
and each perspective includes one to two metrics in brackets: requester (``label quantity'' and ``label quality''), worker (``worker quality detection'' and ``spammer prevention'') and platform (``money cost'').
Except ``worker quality detection'',
other metrics have been popularly used by previous works \citep{shah2015double,shah2016noopps}. They are detailed as follows.

\begin{itemize}

\item Label quantity: we evaluate the label quantity by the percentage of the completion of three tasks. In the skip-stage setting, worker yields unlabeled (uncompleted) data by skipping unsure questions. In the single-stage and the hybrid-stage settings, for objective questions, worker yields (few) unlabeled data because he/she forgets or ignores few questions. For subjective questions, worker yields (more) unlabeled data by inputting invalid answers. For example, they write sentences, such as ``I do not know" in the answer box.
\item Label quality: we evaluate the label quality from two aspects:
(i) the percentage of correct answers and incorrect answers on three tasks;
and (ii) the error of aggregated labels~\citep{shah2015double}. For the $i$th question where $i \in \{1,\ldots, n\}$,
if there are $m_i$ options after majority voting (the tie situation),
and the ground-truth label is one of $m_i$ options,
then we consider that $\frac{1}{m_i}$ of the $i$th question is correct.
Therefore, the error of aggregated labels is $1- (\sum_{i=1}^{n} 1/m_i)/n$.
Since text answers cannot be majority voted on \textit{Speech Clips},
we do not report the error of aggregated labels on \textit{Speech Clips}.

\item Worker quality detection: we evaluate the worker quality detection of the hint-guided approach implicitly, by the error rate (in $\%$) of aggregating original and rescaled crowdsourced labels. For example,
\textit{Sydney Bridge} (origin) means the original labels collected by our approach. For \textit{Sydney Bridge} (rescale), we rank the worker quality from high to low by the usage frequency of the hints in the collection of original labels. Then, we rescale original labels by adaptive weights. Labels from top $20\%$ (bottom $20\%$) workers have been empirically rescaled by $1.8$ ($0.2$). The remaining labels keep unchanged.
If the error rate on rescaled dataset decreases, then we speculate that our hint-guided approach indeed detects the worker quality. Namely, the less usage of hints indicates the higher quality of the worker.

\item Spammer prevention and money cost:
we evaluate the spammer prevention and the money cost by the average payment to each worker. Note that the payment consists of two parts: fixed payment and reward payment. Reward payment is based on a worker's responses to $G$ gold standard questions.
\end{itemize}

\subsection{Experimental Results}

We demonstrate the effectiveness of our hint-guided approach from the following five aspects. Specifically, Section~\ref{sec:labelquan} verifies whether our approach provides a sufficient number of labels. Section~\ref{sec:labelqual} displays whether our approach provides high-quality labels. Section~\ref{exp:worker-quality-detection} denotes whether our approach can detect worker quality. Section~\ref{spammer-prevention} indicates whether our approach prevents spammers. Section~\ref{money-cost} demonstrates whether our approach saves money.

\subsubsection{Label Quantity}
\label{sec:labelquan}
Table \ref{datasets} denotes the percentage of the completion of three tasks.
The first two tasks (\textit{Sydney Bridge} and \textit{Stanford Dogs}) belong to objective questions, while the last task (\textit{Speech Clips}) belongs to subjective questions. Objective questions can be answered by the random guess. Therefore, the percentage of the completion for objective questions is much higher than that for subjective questions. In addition, the hint-guided approach has a high percentage of the completion of both objective and subjective questions. Our approach inspires workers to finish the questions, ensuring the quantity of crowdsourced labels.

\begin{table}[ht]
\caption{Evaluation of the label quantity. We provide the percentage of the completion on three tasks.}
\label{datasets}
\centering
\begin{tabular}{c | c | C{40px} | C{40px}}
	\hline
	\multicolumn{1}{c|}{Data set} & {Baseline}        & {Skip-based} & {\bf Hint-guided} \\ \hline
	   \textit{Sydney Bridge}     & \textbf{100.00}\% & 74.00\%      & 99.11\%           \\ \hline
	   \textit{Stanford Dogs}     & 99.72\%           & 58.18\%      & \textbf{99.91}\%  \\ \hline
	    \textit{Speech Clips}     & 58.33\%           & 30.00\%      & \textbf{75.00}\%  \\ \hline
\end{tabular}
\end{table}

\subsubsection{Label Quality}
\label{sec:labelqual}
\begin{figure*}[ht]

\centering
\subfigure[\textit{Sydney Bridge}.]
{
	\centering
	\includegraphics[scale=0.35]{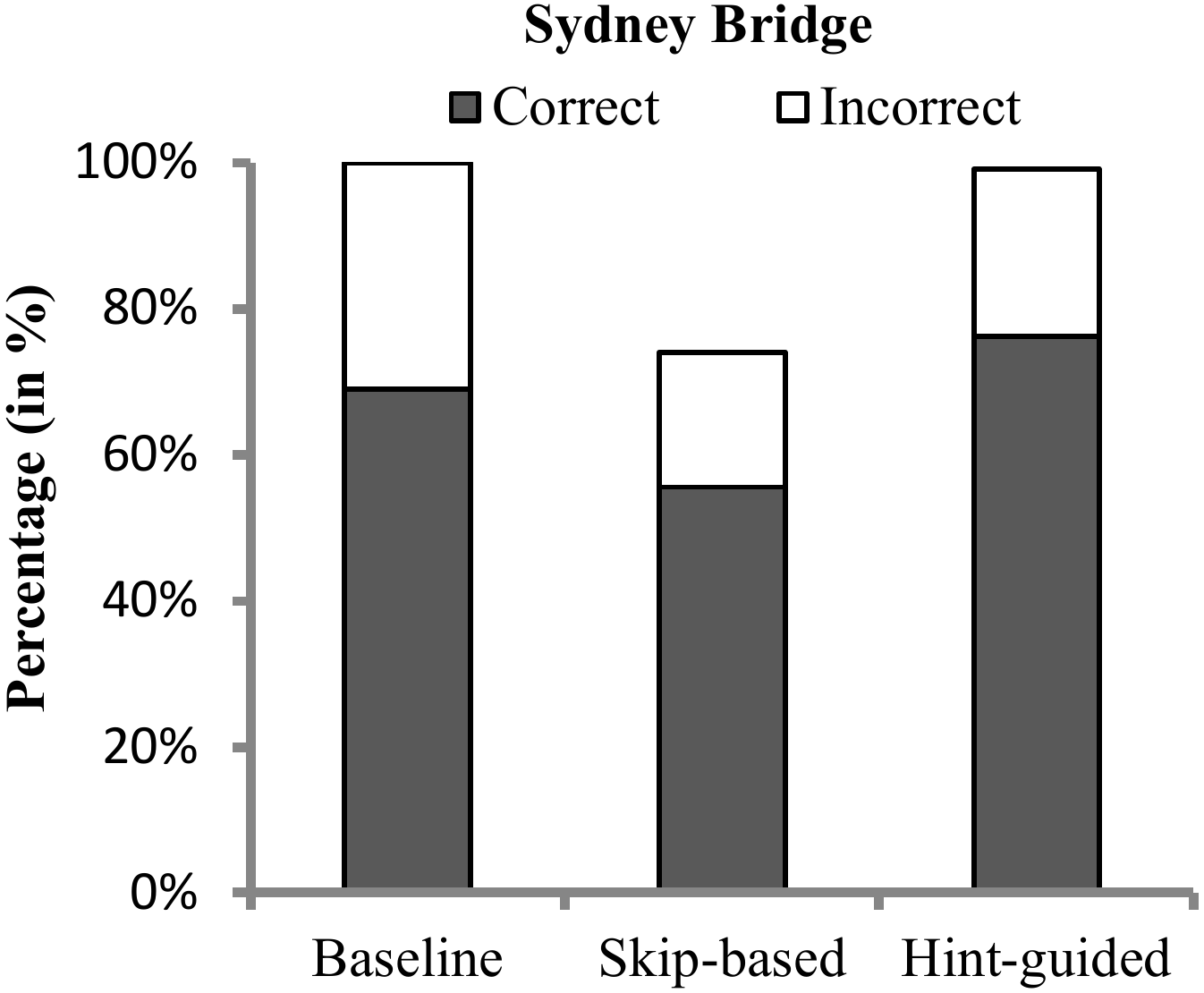}
	\label{per-Sydney}
}
\subfigure[\textit{Stanford Dogs}.]
{
	\centering
	\includegraphics[scale=0.35]{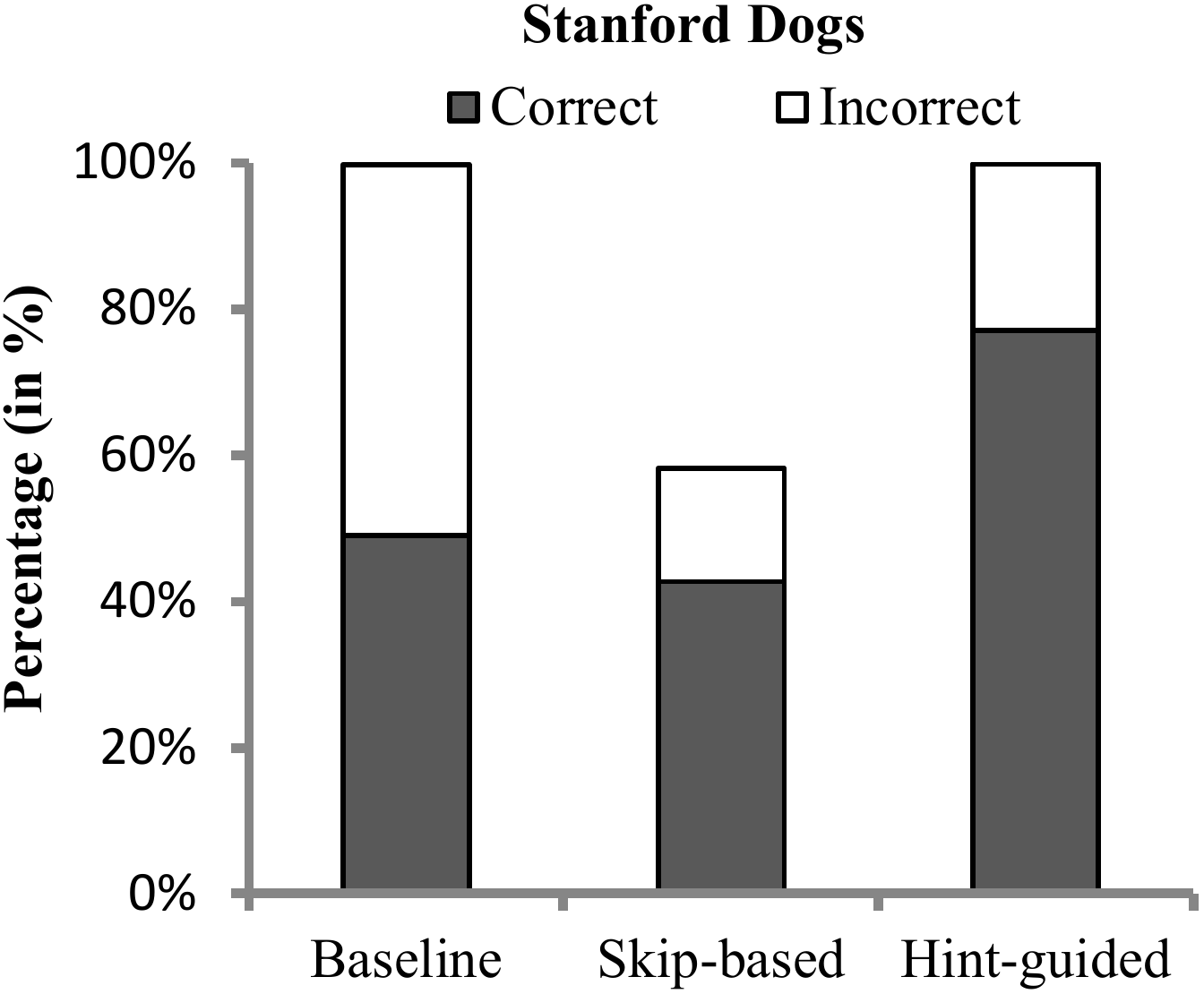}
	\label{per-Stanford}
}
\subfigure[\textit{Speech Clips}.]
{
	\centering
	\includegraphics[scale=0.35]{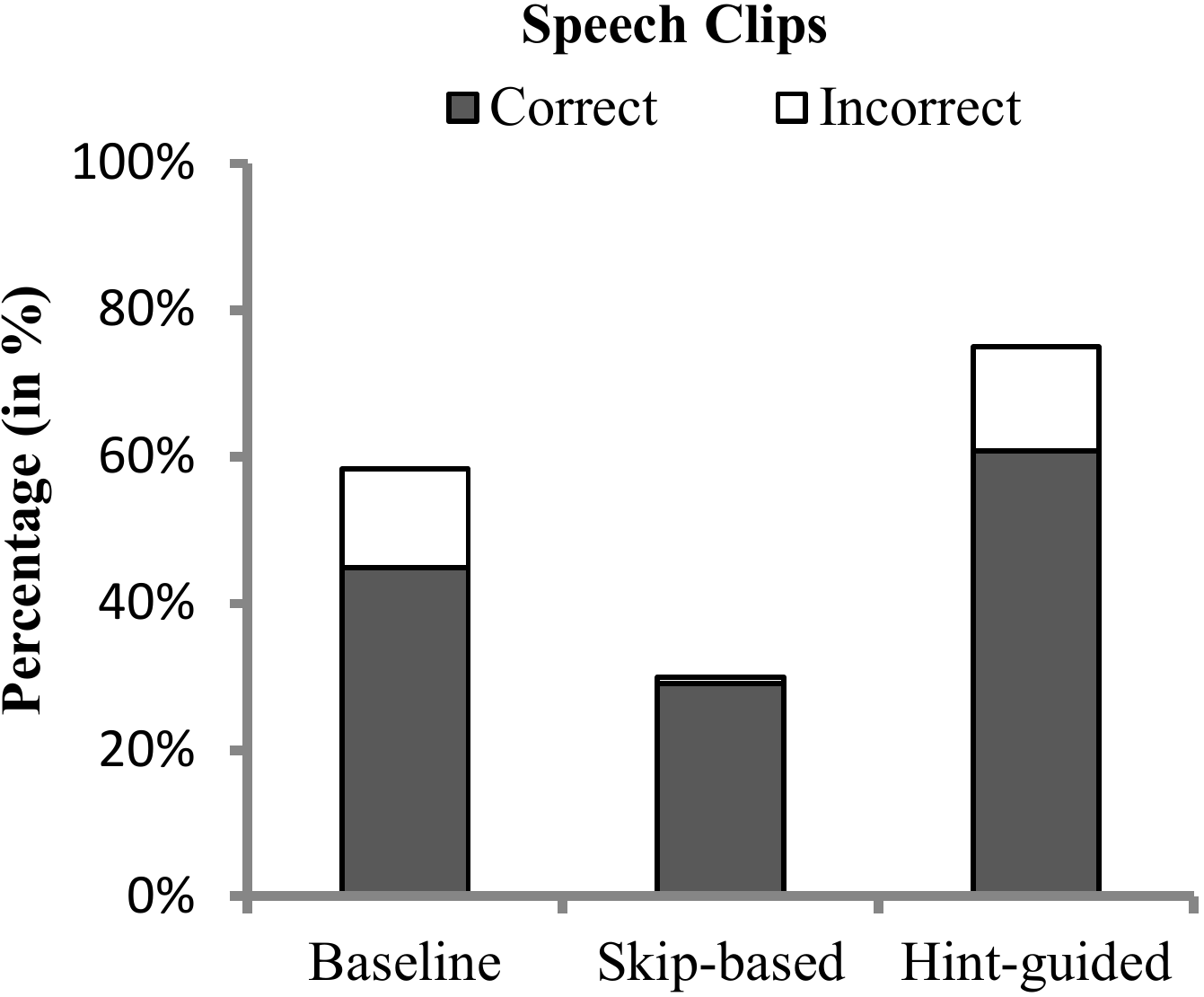}
	\label{per-Speech}
}

\caption{\label{real-experiment-1} Evaluation of label quality.
	Percentage (in \%) of correct answers and incorrect answers on three tasks are provided. Note that, we do not plot the percentage of unlabeled questions.
}
\end{figure*}

Figure \ref{real-experiment-1} plots the percentage of correct answers and incorrect answers on three tasks.
First, on all tasks, the percentage of correct answers in the hint-guided approach is higher than that in the baseline and skip-based approaches. Second, on \textit{Speech Clips}, the percentage of incorrect answers is extremely low in the skip-based approach. The reason is that most people skip difficult speech clips, and answer several easy ones. Third, compared with other approaches, our hint-guided approach ensures a sufficient number of high-quality labels.

\begin{figure}[ht]
	\centering
	\subfigure[\textit{Sydney Bridge}.]
	{
		\centering
		\includegraphics[scale=0.35]{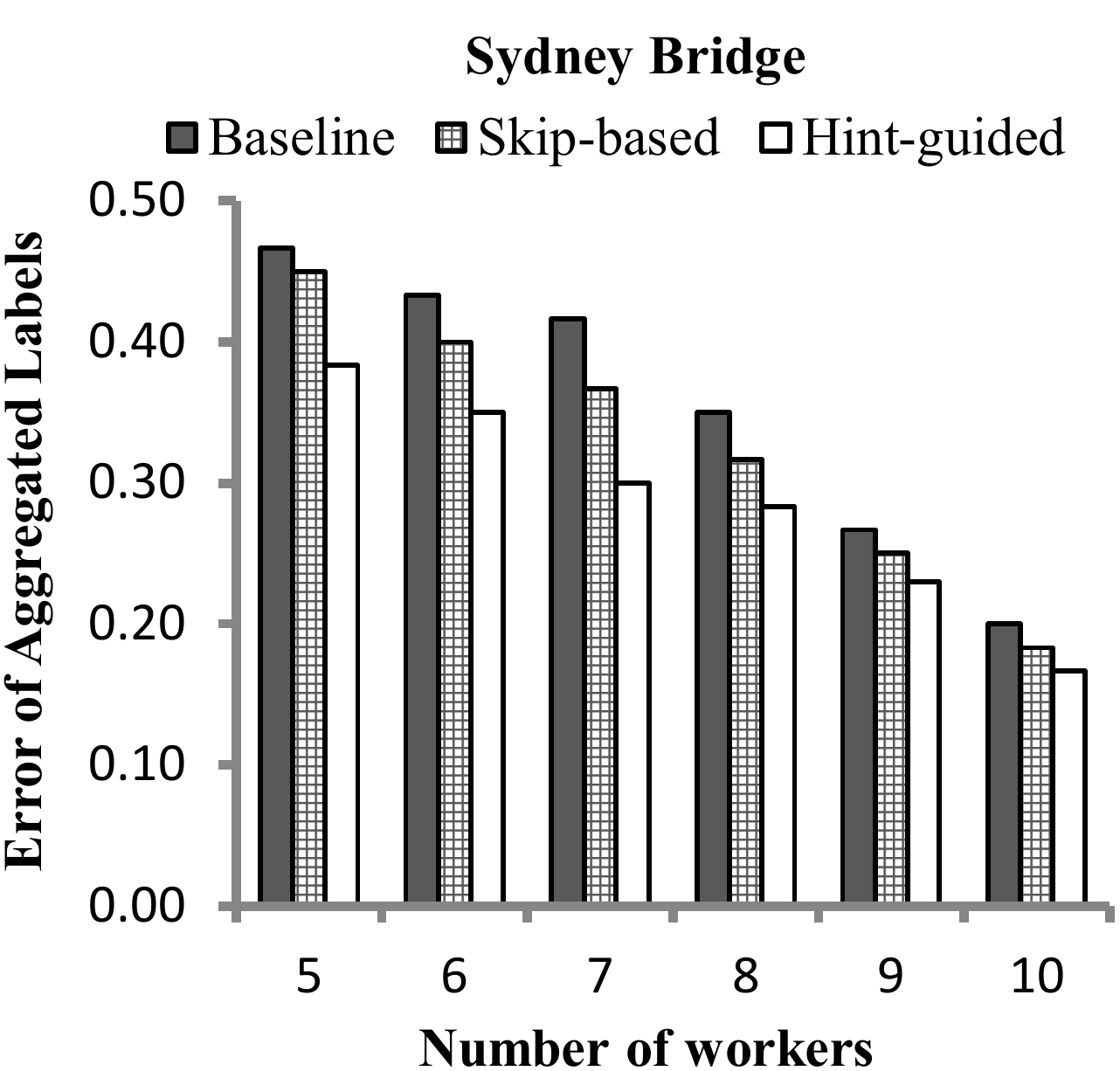}
		\label{err-Sydney-Bridge}
	}
	\subfigure[\textit{Stanford Dogs}.]
	{
		\centering
		\includegraphics[scale=0.35]{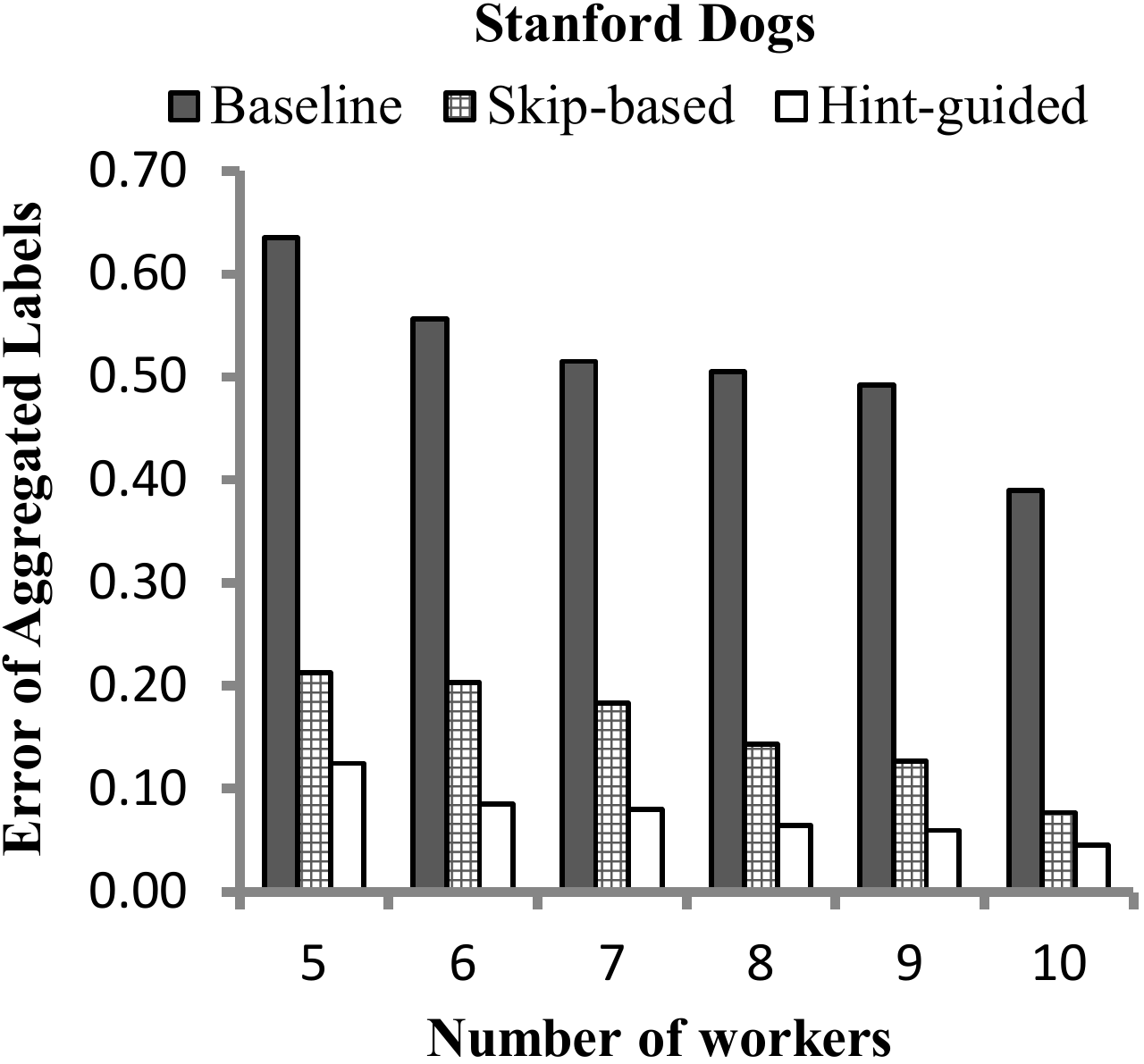}
		\label{err-Stanford-Dogs}
	}
	
	\caption{Evaluation of the label quality.
		Results on \textit{Speech Clips} are not reported,
		as text answers cannot be majority voted.}
	\label{real-experiment-2}
	
\end{figure}

Figures \ref{err-Sydney-Bridge} and \ref{err-Stanford-Dogs} plot the error of aggregated labels on
the \textit{Sydney Bridge} and \textit{Standford Dogs} tasks. The number of workers (abbreviated as \emph{n\_workers}) is set to
$\{5,6,7,8,9,10\}$, since the error of aggregated labels comes from majority voting among multiple workers \citep{shah2015double}, and the number of multiple
workers depends on varying situations. For each of combinations between tasks and \emph{n\_workers}, we perform the following actions $200$ times repeatedly. In each time, for all questions, we randomly select \emph{n\_workers} workers and perform the majority voting on their responses to yield the aggregated labels. The plotted error of aggregated labels is averaged across $200$ results. We observe that the hint-guided approach consistently outperforms the baseline and the skip-based approaches, and the performance gap between the baseline and the hint-guided approaches is extremely obvious on \textit{Stanford Dogs}.

\subsubsection{Worker Quality Dectection}
\label{exp:worker-quality-detection}

Table~\ref{worker-quality-detection} denotes the error of aggregating original and rescaled crowdsourced labels.
For rescaled crowdsourced labels, labels from estimated high-quality workers are adaptively given more weights, and vice versa. From Table~\ref{worker-quality-detection}, we can see the error of aggregating rescaled labels is lower than the error of aggregating original labels. It demonstrates that our hint-guided approach can detect the high-quality workers effectively.
Then, the error decreases significantly on
\textit{Sydney Bridge}, since the size of \textit{Sydney Bridge} is relatively small ($30$ questions) compared to \textit{Stanford Dogs} ($100$ questions). We believe that, the informative hints for \textit{Stanford Dogs} may guide the low-quality workers to make more accurate decisions. Then, the performance gap between high-quality and low-quality workers is insignificant.
Therefore, the effect of label rescaling is marginal on this dataset.

\begin{table}[H]
\centering
\caption{Evaluation of the worker quality detection of the hint-guided approach.
	Error rate (in \%) is provided for aggregating original and rescaled crowdsourced labels. }

\begin{tabular}{c | c | C{50px} | C{50px}}
	\hline
	\multicolumn{2}{c|}{Number of Workers} & 5           & 10             \\ \hline
	\textit{Sydney Bridge} & origin        & 38.33\%       & 16.67\%          \\ \cline{2-4}
	                       & rescale       & \textbf{30.00\%} & \textbf{11.67\%} \\ \hline
	\textit{Stanford Dogs} & origin        & 12.50\%        & 4.50\%            \\ \cline{2-4}
	                       & rescale       & \textbf{12.00\%} & \textbf{4.00\%}     \\ \hline
\end{tabular}
\label{worker-quality-detection}
\end{table}

\subsubsection{Spammer Prevention}\label{spammer-prevention}
\begin{figure}[!tp]
\centering
\subfigure[]
{
   \includegraphics[scale=0.35]{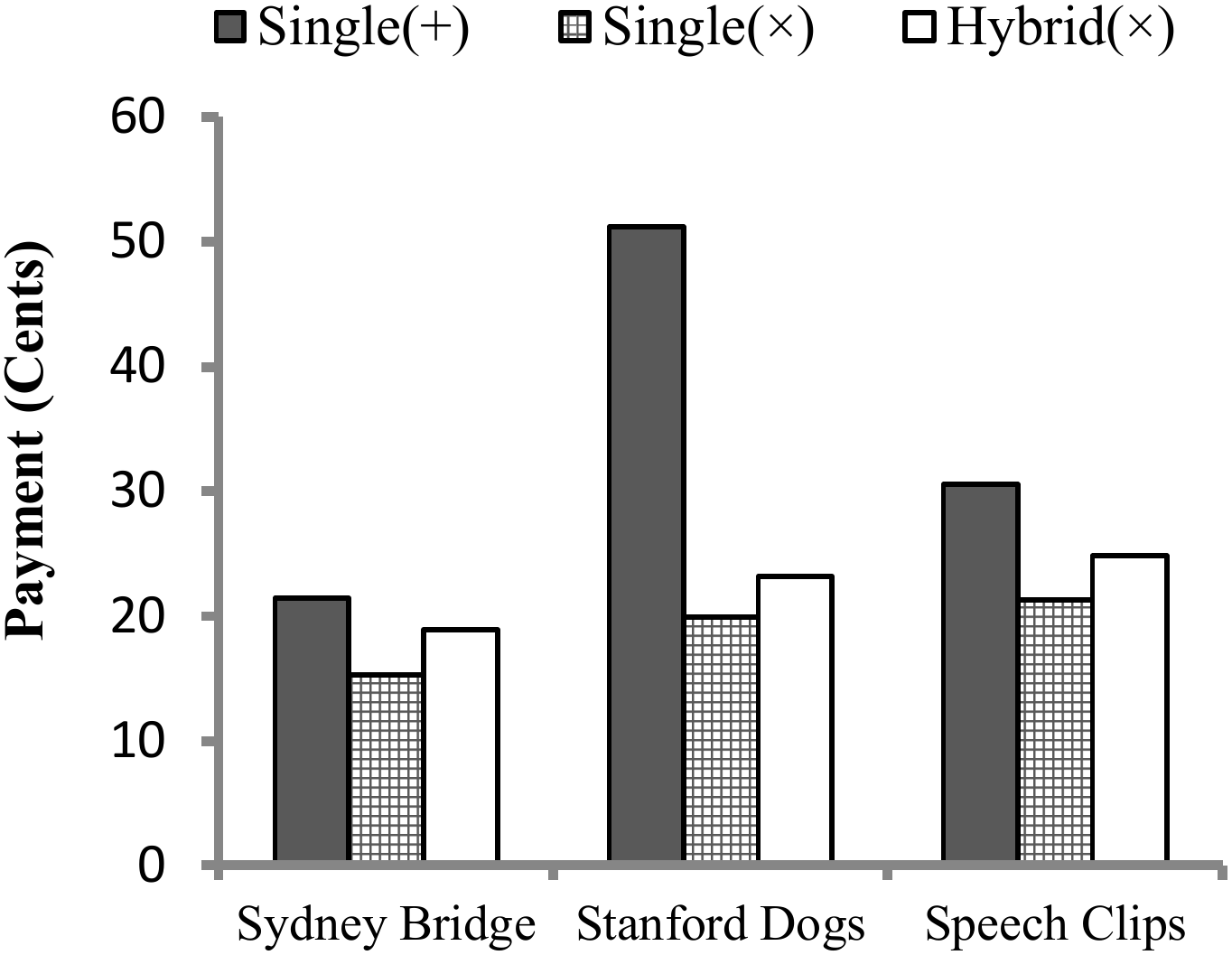}
   \label{averaged-payment-1}
}
\subfigure[]
{
   \includegraphics[scale=0.35]{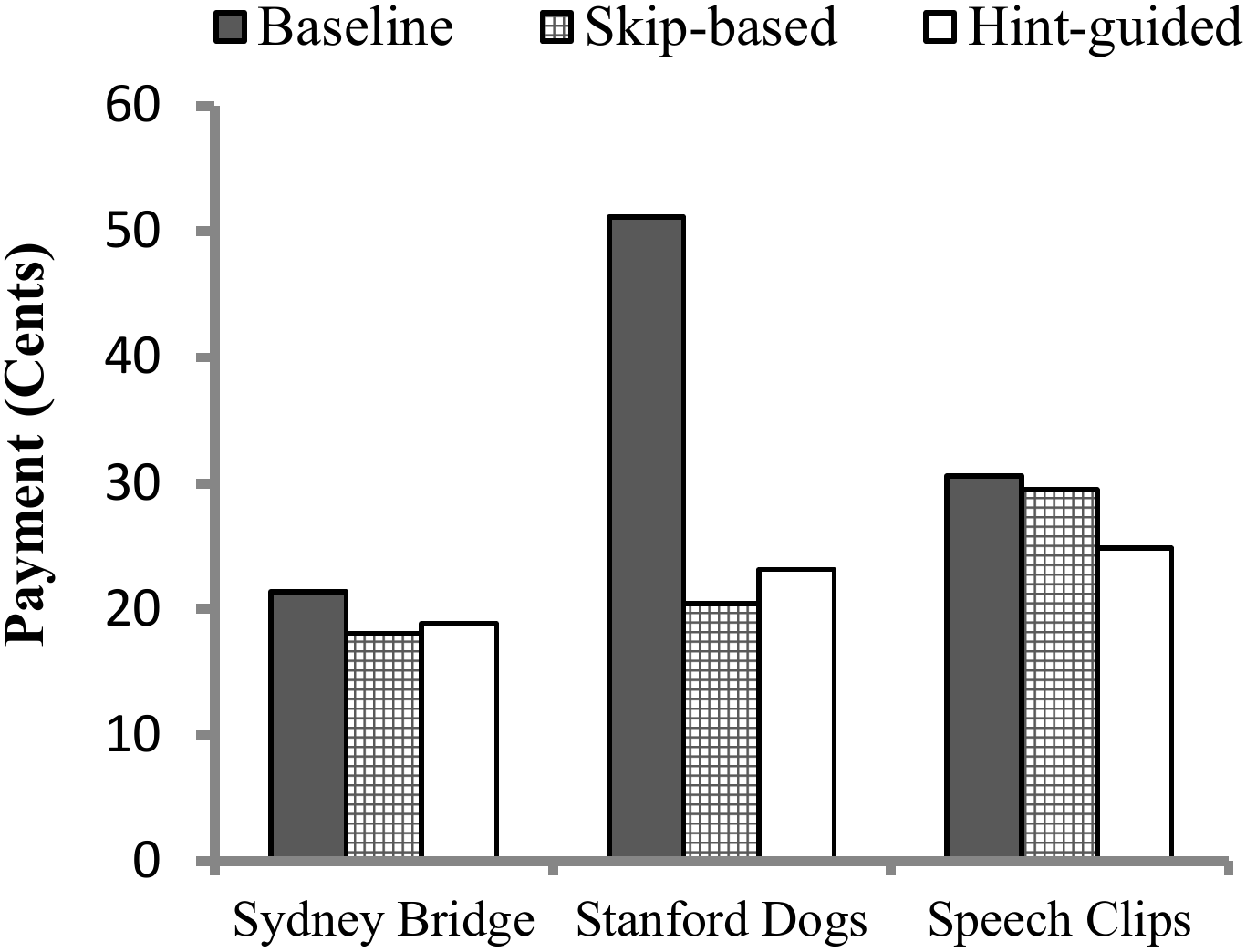}
   \label{averaged-payment-2}
}
\caption{Evaluation of the spammer prevention.
Average payment to each worker on all three tasks are provided. Evaluation of the money cost.}
\end{figure}

The baseline and hint-guided approaches are represented as Single($+$) and Hybrid($\times$), respectively.
We provide one extra interaction: the single-stage setting with the ``$\times$" mechanism (Single($\times$)), and all parameters are consistent. Figure \ref{averaged-payment-1} explores how our approach prevents spammers. It plots the average payment to each worker under three approaches.
We have one observation: the payments of Single($\times$) and Hybrid($\times$) are lower than that of Single($+$), since an answer in $G$ questions is incorrect, and thus the reward of the ``$\times$" mechanism becomes zero. Since spammers answer each question randomly, the ``$\times$" mechanism used by our approach makes the smallest payment to them. Thus, our approach prevents spammers.

\subsubsection{Money Cost}\label{money-cost}
Figure~\ref{averaged-payment-2} plots the average payment to each worker under the three approaches. The higher the payment is, the worse the economy of the approach. The payment is calculated as the average of the payments across $200$ random selections of $G$ questions. This process mitigates the distortion of results caused by the randomness in the choice of $G$ questions.
We can see that, the payments of the skip-based and hint-guided approaches are comparable but less than the payment of the baseline approach, especially in the \textit{Stanford Dogs} task, since both the skip-based and hint-guided approaches use the multiplicative mechanism but the baseline approach use the additive mechanism. Thus, from the perspective of saving money, we should not employ the baseline approach. Note that, on the \textit{Sydney Bridge} and \textit{Stanford Dogs} tasks, although the payment in the skip-based approach is slightly lower than that in the hint-guided approach, the number of high-quality labels from the hint-guided approach is obviously higher than that from the skip-based approach
(Figure~\ref{real-experiment-2}).

\section{Conclusions}\label{Conclusions}
To improve the label quality, we proposed a hint-guided approach that encourages workers to use hints when they answer unsure questions. Our approach consists of the hybrid-stage setting and the hint-guided payment mechanism. We proved the incentive compatibility and uniqueness of our mechanism. Besides, our approach can detect the high-quality workers for more accurate result aggregation. Comprehensive experiments conducted on Amazon MTurk revealed the effectiveness of our approach and validated the simple and practical deployment of our approach. These merits are critical for the success of many machine learning applications in practice.

As for future works, first, the hint-guided approach is designed under the worker's independence.
However, it would become more interesting to extend hint-guided approach under the worker's dependence, where the reward of a worker depends on the answers of the other ones.
Second, we hope to extend the hybrid-stage setting from binary choice to multiple choice with the corresponding theoretical results.
Third, we consider to provide hints from different levels for all questions. Specifically, we will provide the hints from coarse to fine, which corresponds the different expected payments.
Finally,
some workers may still be very confused even with hints,
we may mix up the unsure option in the hint stage to further improve the label quality further.


\bibliographystyle{spbasic}      
\bibliography{example_paper}

\end{document}